\documentclass[aps,prl,amsmath,amssymb,letterpaper,showpacs,reprint,floatfix,superscriptaddress]{revtex4-1}
\usepackage{placeins,color}
\usepackage[caption=false]{subfig}

\usepackage[allcolors=blue,colorlinks=true,breaklinks=true,bookmarks=false]{hyperref}

\usepackage{graphicx}

\begin{document}
\newcommand{\changed}[1]{{\textcolor{green}{#1}}}
\newcommand{\missing}[1]{{\textcolor{red}{MISSING: #1}}}

\newcommand{\ifw}[0]{\affiliation{Leibniz Institute for Solid State and Materials Research (IFW) Dresden, D--01069 Dresden , Germany}}
\newcommand{\ifp}[0]{\affiliation{Institute for Solid State Physics, Dresden University of Technology, D--01069 Dresden, Germany}}
\newcommand{\lmu}[0]{\affiliation{Laboratory for Muon--Spin Spectroscopy, Paul Scherrer Institute, CH--5232 Villigen PSI, Switzerland}}
\newcommand{\kip}[0]{\affiliation{Kirchhoff Institute for Physics, Heidelberg University, D--69120 Heidelberg, Germany}}
\newcommand{\bessy}[0]{\affiliation{Helmholtz--Zentrum Berlin GmbH für Materialien und Energie, BESSY, D--12489 Berlin, Germany}}

\newcommand{\mpi}[0]{\affiliation{Max Planck Institute for the Physics of Complex Systems, D--01187 Dresden, Germany}}
\newcommand{\bochum}[0]{\affiliation{Institute for Theoretical Physics III, Ruhr--University Bochum, D--44801 Bochum, Germany}}

\newcommand{\ud}[0]{Ba$_{0.75}$Na$_{0.25}$Fe$_2$As$_2$}
\newcommand{\co}[0]{Ba$_{0.70}$Na$_{0.30}$Fe$_2$As$_2$}
\newcommand{\od}[0]{Ba$_{0.60}$Na$_{0.40}$Fe$_2$As$_2$}
\newcommand{\ba}[0]{Ba$_{1-x}$Na$_{x}$Fe$_2$As$_2$}
\newcommand{\bk}[0]{Ba$_{1-x}$K$_{x}$Fe$_2$As$_2$}
\newcommand{\heading}[1]{}


\author{Hemke Maeter} \email{h.maeter@physik.tu-dresden.de} \ifp{}
\author{Gwendolyne Pascua} \lmu{}
\author{Hubertus Luetkens} \lmu{}
\author{Johannes Knolle} \mpi{}
\author{Saicharan Aswartham} \ifw{}

\author{Sabine Wurmehl} \ifw{} \ifp{}
\author{G\"unter Behr} \ifw{}
\author{Bernd B\"uchner} \ifw{} \ifp{}

\author{Zurab Shermadini} \lmu{}
\author{Kamil Sedlak} \lmu{}
\author{Alex Amato} \lmu{}

\author{Roderich Moessner} \mpi{}
\author{Ilya Eremin} \bochum{}

\author{Hans-Henning Klauss} \ifp{}


\title{Strong Competition of Superconducting and Magnetic Order Parameters in Ba$_{1-x}$Na$_x$Fe$_2$As$_2$}

\begin{abstract}
We study the interplay of magnetic and superconducting order in single crystalline hole doped \ba{} using muon spin relaxation.
We find microscopic coexistence of magnetic order and superconductivity.
In a strongly underdoped specimen the two forms of order coexist without any measurable reduction of the ordered magnetic moment by superconductivity, while in a nearly optimally doped sample the ordered magnetic moment is strongly suppressed below the superconducting transition temperature.
This coupling can be well described within the framework of an effective two-band model incorporating inter- and intra-band interactions.
In optimally doped \ba{} we observe no traces of static or dynamic magnetism and the temperature dependence of the superfluid density is consistent with two s-wave gaps without nodes.
\end{abstract}
\maketitle

Strongly correlated electron systems like the cuprate high-$T_c$, organic, and heavy-fermion superconductors exhibit a delicate interplay of superconductivity and antiferromagnetic order.
Heavy-fermion superconductivity often takes place in the vicinity of a magnetic quantum critical point \cite{mathur98}.
Similarly, magnetic order lies nearby and competes with superconductivity in the phase diagrams of the cuprate high-$T_c$ superconductors\cite{cuprates1,cuprates2,cuprates3}.
In the ferropnictide superconductors the interplay of magnetism and superconductivity leads to a new phase that is both homogeneously superconducting and magnetically ordered \cite{parker09,fernandes10,vorontsov10,fernandes10,fernandes10b}---i.e., magnetic order and superconductivity coexist microscopically \cite{pratt09,fernandes10,mazin10}.
Still, both instabilities compete for the same electron states around the Fermi surface, which causes a reduction of the ordered magnetic moment upon cooling below $T_c$.
In some materials electronic phase separation appears instead of microscopic coexistence.
Experiments with electron doped BaFe$_{2-x}$Co$_x$As$_2$ suggest microscopic coexistence \cite{pratt09,julien09,nandi10}, nevertheless a recent report by \citet{bernhard12} indicates electronic inhomogeneity are important near optimal doping.
Electronic phase separation, on the other hand, commonly occurs in hole doped \bk{} \cite{park09,aczel08,goko09,julien09}.
Only recently \citet{wiesenmayer11} showed a reduction of the ordered moment by superconductivity in polycrystalline underdoped \bk{}---an unambiguous proof of microscopic coexistence---consistent with other reports \cite{rotter09,urbano10,avci11}.

In this Letter we report a strong interplay of the order parameters of superconductivity and magnetic order in \ba{}.
In a nearly optimally doped compound, the ordered magnetic moment of the Fe spin density wave order, that appears below $T_N\approx 40$~K, is reduced by $\approx 65$\% upon cooling below $T_c\approx 30$~K.

\heading{Experimental}
We examined platelike Ba$_{1-x}$Na$_x$Fe$_2$As$_2$ single crystals (growth conditions and characterization are described in Ref.~\cite{aswartham12}) with the $c$-axis oriented along the muon beam.
The muon-spin relaxation was measured in zero (ZF) and transverse (TF) magnetic fields of up to 0.64~T using the DOLLY and GPS instruments at the Paul Scherrer Institute equipped with $^4$He flow-cryostats.
To analyze the $\mu$SR data, we used the {\sc musrfit} framework \cite{musrfit}.
The stoichiometry of each crystal was determined with electron-dispersive x-ray spectroscopy (EDX).
Magnetization, electrical resistivity, angle-resolved photoemission spectroscopy (ARPES), Hall-effect, and specific-heat measurements of single crystals from the same batches are published in Refs.~\cite{pramanik11,aswartham12,pramanik12}.


\heading{ZF-$\mu$SR} Fig.~\ref{img.zf} shows the muon spin polarization $P(t)$ in ZF perpendicular to the beam.
For $x=0.40$, the relaxation of $P(t)$ is temperature independent and characteristic for a paramagnetic material (down to the lowest measured temperature of 2~K).
For $x=0.25$, and 0.30 we find two precession signals $\alpha=A, B$ with frequencies $f_\alpha$ that indicate two magnetically nonequivalent muon sites, and prove long range magnetic order with $T_N=123(1)$, and 40.0(5)~K, respectively.
We analyze $P(t)$ with two damped-cosine oscillations, and a nonrelaxing signal due to paramagnetic and non-oscillating signals. 

\begin{figure*}[htb]
\captionsetup[subfloat]{position=top,parskip=0pt,aboveskip=0pt,justification=raggedright,labelseparator=none,farskip=0pt,nearskip=2pt,margin=0pt,captionskip=-15pt,font=normalsize,singlelinecheck=false}
\subfloat[]{\includegraphics[width=0.99\textwidth]{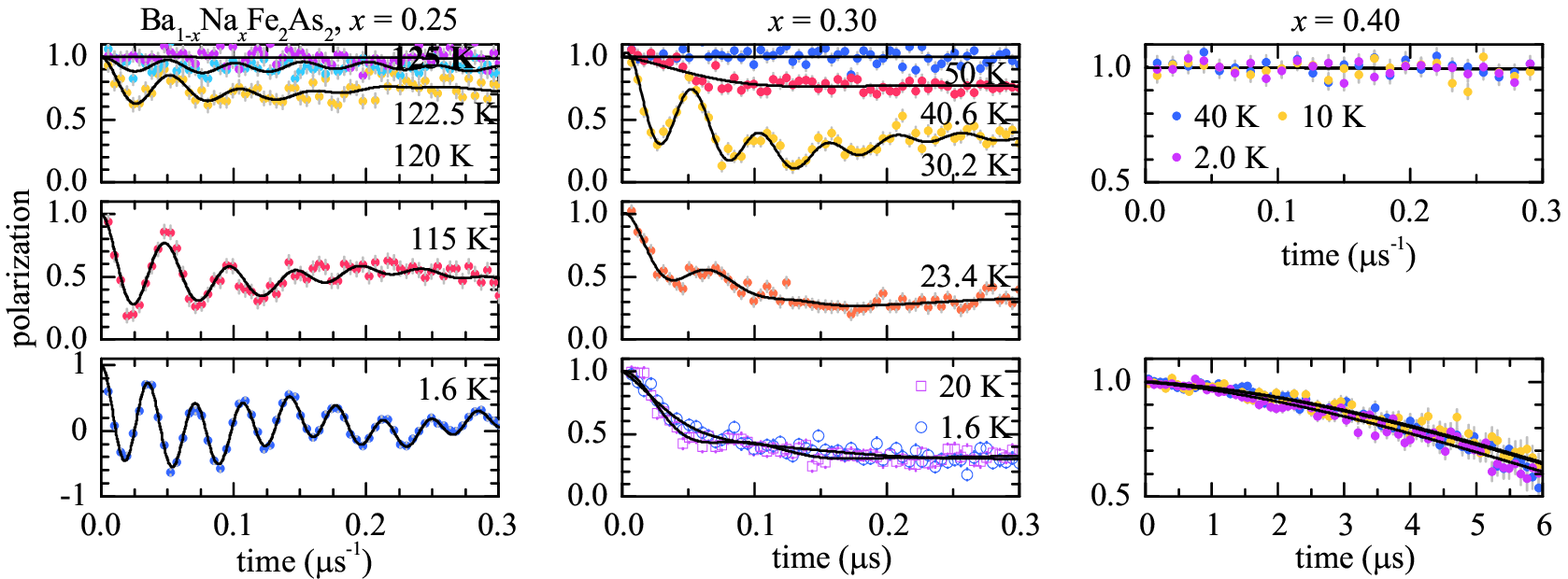} \label{img.zf}}\\
\subfloat[]{\includegraphics[width= 8.5cm]{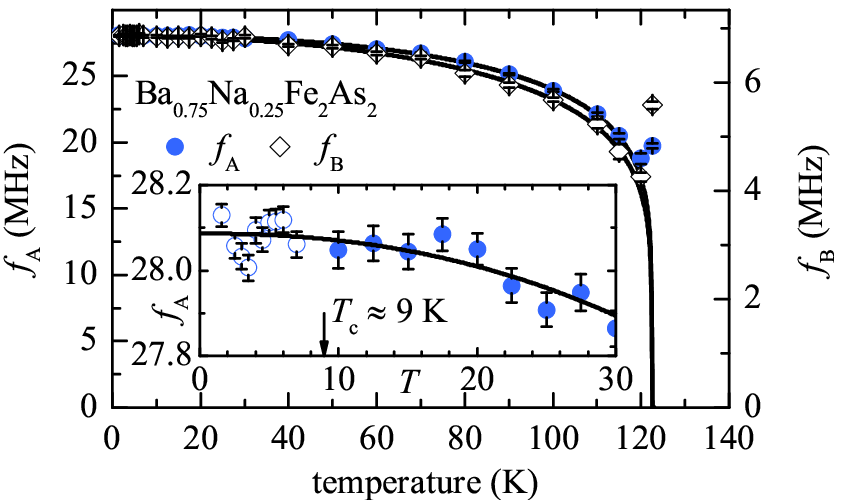} \label{img.25-freq}}\quad
\subfloat[]{\includegraphics[width= 8.5cm]{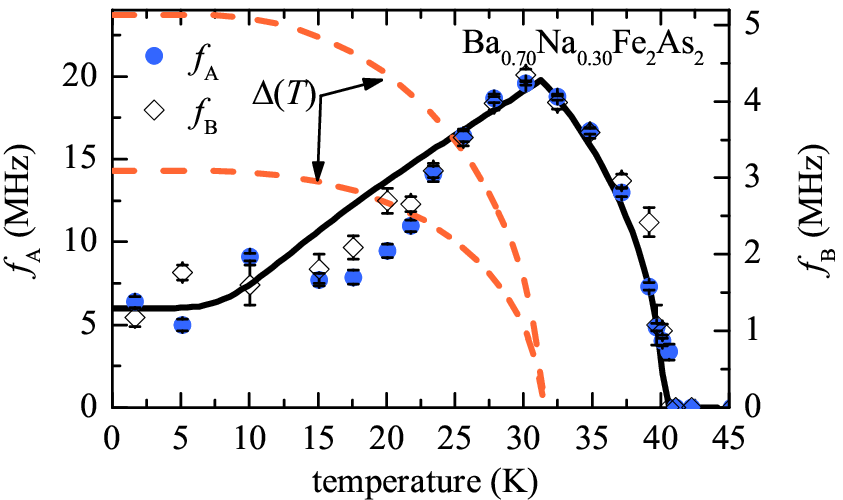} \label{img.35-freq}}
\caption{ZF $\mu$SR of \ba{}: (a) the muon spin polarization time dependence for $x=0.25$ (left), 0.30 (middle), and 0.40 (right); the temperature dependence of the magnetic order parameter ($\propto f_\alpha(T)$) for (b) $x=0.25$ does not show any anomaly at $T_c$, the inset shows the low temperature regime in which superconductivity occurred (the open symbols indicate the data \emph{excluded} from the fit; the lines are fits, see text), but for (c) $x=0.30$, the precession frequencies (the magnetic order parameter) are reduced by $\approx 65$\% below superconducting critical temperature $T_{c2}$ (the line is calculated from an itinerant two-band model, see text)---nevertheless, all evidence supports microscopic coexistence in both cases.}
\label{img.zf-all}
\end{figure*}

\heading{Magnetic order parameter}
The muon-spin precession frequencies (which are proportional to the magnetic order parameter) of $x=0.25$, and 0.30 are shown in Figs.~\ref{img.25-freq}, and \ref{img.35-freq}.
For $x=0.25$ we find no anomaly $f_\alpha(T)$ at $T_c\approx 9$~K on cooling ($T_c$ from Ref.~\cite{aswartham12}).
A fit with a general order parameter temperature dependence $f_\alpha(T)=f_\alpha(0)(1-(T/T_N)^\alpha)^\beta$, shown in Fig.~\ref{img.25-freq}, to the data \emph{above} $T_c$ reproduces the data also \emph{below} $T_c$.
By setting $\alpha=1$ we estimate $\beta=0.13(1)$, consistent with the 2D-Ising universality class ($\beta=0.125$), and similar to results for BaFe$_2$As$_2$ \cite{wilson}.

We conducted $\mu$SR ``pinning experiments'' to verify bulk superconductivity:
We cooled the sample with $x=0.25$ to $T=5.0$~K in a transverse magnetic field $\mu_0 H_1 = 500$~mT, recorded a time histogram (Fourier transform shown in Fig.~\ref{img.25-pinning} as blue line), then isothermally reduced the field  to $\mu_0 H_2=450$~mT and recorded another histogram (red line).
The magnetic order caused four precession signals in the external magnetic field: $\gamma_\mu\mu_0 H_1\pm 2\pi f_A$, and $\gamma_\mu\mu_0 H_1\pm 2\pi f_B$, as indicated by the Fourier transform in Fig.~\ref{img.25-pinning}, in addition to the background signal due to muons stopped by the cryostat walls or the detector.
The signal fraction that followed the change of the external magnetic field was $\approx 10$\% of the total signal, which is consistent with a background signal.
The remainder of the signal maintained its average internal field after we reduced the field, which indicates strong flux pinning due to superconductivity in the majority of the sample.
Instead, caused by enhanced vortex lattice disorder, only the damping of the spin precession increased (spectral lines broadened).
This evidence for bulk superconductivity taken together with the bulk magnetic order supports microscopic coexistence of magnetic order and superconductivity, albeit with a coupling of the order parameters that is too small to be detectable \cite{fernandes10-order}.

The pinning experiment for $x=0.30$, shown in Fig.~\ref{img.35-pinning}, also indicates bulk superconductivity.
In contrast to $x=0.25$, the magnetic order parameter for $x=0.30$ (see Fig.~\ref{img.35-freq}) was suppressed by $\approx 65$\% on cooling below $T_{c2}=29.0(6)$~K, which proves microscopic coexistence of magnetic order and superconductivity and strong coupling of both order parameters.

\begin{figure}[htb]
\centering
\captionsetup[subfloat]{position=top,parskip=0pt,aboveskip=0pt,justification=raggedright,labelseparator=none,farskip=0pt,nearskip=2pt,margin=0pt,captionskip=-10pt,font=normalsize,singlelinecheck=false}
\subfloat[]{\includegraphics[width= 5.8cm]{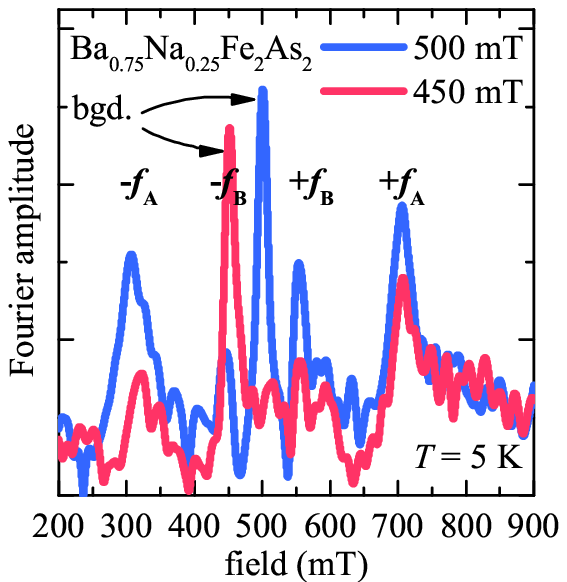} \label{img.25-pinning}}\\
\subfloat[]{\includegraphics[width= 5.8cm]{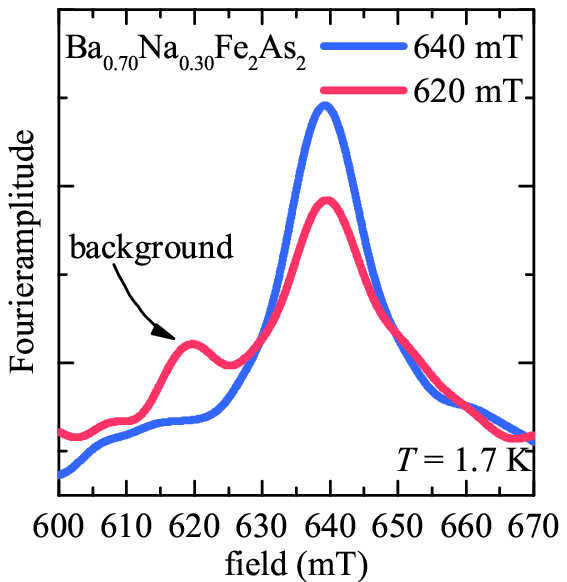} \label{img.35-pinning}}
\caption{Bulk superconductivity is indicated by TF-$\mu$SR pinning experiments for (a) $x=0.25$:
After isothermally changing the field only $\approx 10$\% of the signal precesses with $f_\mu\approx\gamma_\mu/(2 \pi)\cdot 450$~mT due to muons that stop in the cryostat walls and the sample holder (bgd.),
$\pm f_\alpha$ indicate signals from muon sites $\alpha=A$, $B$ with the local field parallel ($+$)/anti-parallel ($-$) to the applied field,
and for (b) $x=0.30$ the background signal amounts to $\approx 5$\% of the total signal---both experiments indicate strong flux pinning due to superconductivity in the majority of the sample volume.}
\label{img.sc-all}
\end{figure}

Theoretical studies show that microscopic coexistence as well as phase separation arise in itinerant multi-band models as a function of Fermi surface topology, band dispersion, and symmetry of the superconducting order parameter \cite{parker09,fernandes10,vorontsov10,fernandes10,fernandes10b}.
Within these models, $s_\pm$ Cooper-pairing, in which the gap changes sign between the hole and electron pockets, has the largest tendency towards microscopic coexistence.
In order to show that such an approach can account for our measurements, we calculate the temperature dependence of the ordered magnetic moment $M(T)$ within a simplified two-band model incorporating inter- and intra-band interactions that drive the spin density wave order (SDW) as well as superconductivity (SC) \cite{brydon09,knolle11}:
$H=\sum_{\mathbf{k} \sigma}\{\epsilon_c(\mathbf{k})c^\dagger_{\mathbf{k} \sigma}c_{\mathbf{k} \sigma}+\epsilon_f(\mathbf{k})f^\dagger_{\mathbf{k} \sigma}f_{\mathbf{k} \sigma}\}+\sum_{\mathbf{k},\mathbf{k}^\prime,\mathbf{q},\sigma,\sigma^\prime} \{ u_1 c^\dagger_{\mathbf{k}+\mathbf{q} \sigma} f^\dagger_{\mathbf{k}^\prime-\mathbf{q} \sigma^\prime} f_{\mathbf{k}^\prime \sigma^\prime} c_{\mathbf{k} \sigma}+ \frac{u_3}{2}(f^\dagger_{\mathbf{k}+\mathbf{q} \sigma} f^\dagger_{\mathbf{k}^\prime-\mathbf{q} \sigma^\prime} c_{\mathbf{k}^\prime \sigma^\prime} c_{\mathbf{k} \sigma}+\textrm{H.c.})\}$, with the dispersion $\epsilon_c(\mathbf{k})$ of the electron ($c$) and $\epsilon_f(\mathbf{k})$ of the hole ($f$) band.
Decoupling the interactions in mean-field we diagonalize the Hamiltonian by sequential Bogolyubov transformations that lead to the eigenenergies
$\Omega_{\mathbf{k}}^{\gamma} = [\left( E_{\mathbf{k}}^{\gamma} \right)^2 + \left( \Delta_{\mathbf{k}}^{\gamma}\right)^2]^{0.5}$ where $\gamma=\alpha,\beta$ and
$E_{\mathbf{k}}^{\alpha,\beta} = \epsilon_{\bf k}^{+} \pm [\left( \epsilon_{\bf k}^{-}\right)^2 + M^2]^{0.5}$. We use the following parameters appropriate for a two-band model with hole doping: $t=1.0, t^\prime=0.72, \epsilon_f=1.3, \epsilon_c=-3.3, \mu=0.1,  u_{SDW}=u_1+u_3=4.985, u_{sc}=u_3=5.605$.
Here $M$ is the antiferromagnetic (SDW) order parameter and $\Delta_{\mathbf{k}}^{\gamma}$ are the two SC gaps (see Ref.~\cite{knolle11} for details).
We solve the self-consistency equations in order to obtain the temperature dependence of the magnetic order parameter $M(T)$, and the gaps $\Delta^\alpha_\mathbf{k}(T)$, $\Delta^\beta_\mathbf{k}(T)$ of $s_\pm$ superconductivity.

The results for $M(T)$ and the superconducting gaps are shown as a solid/dashed lines in Fig.~\ref{img.35-freq}. The dip in $M(T)$ coincides with the onset of SC at $T_c$. In particular, the relative reduction of the magnetic order parameter below $T_c$ is mostly determined by the magnitude of the superconducting order parameter that competes for the same Fermi surface points. More elliptical electron pockets reduce the magnitude of $M(T)$ in comparison to the superconducting order parameter, thus increasing the relative suppression.
The agreement between the calculated temperature dependence and the experimental data is very good for $T>25$~K.
At lower temperatures the experimental data drop somewhat faster than the theoretical temperature dependence.
The discrepancy may be connected to a change of magnetic fluctuations that change the magnetic moment at the SC $T_c$ due to gapping of particle-hole excitations.

\begin{figure}[htb]
\captionsetup[subfloat]{position=top,parskip=0pt,aboveskip=0pt,justification=raggedright,labelseparator=none,farskip=0pt,nearskip=0pt,margin=0pt,captionskip=-10pt,font=normalsize,singlelinecheck=false}
\subfloat[]{\includegraphics[width=8.5cm]{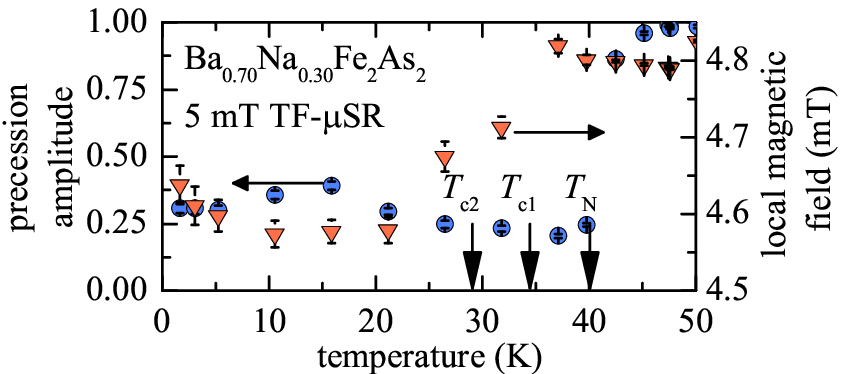} \label{img.tf}}\\
\subfloat[]{\includegraphics[width=8.5cm]{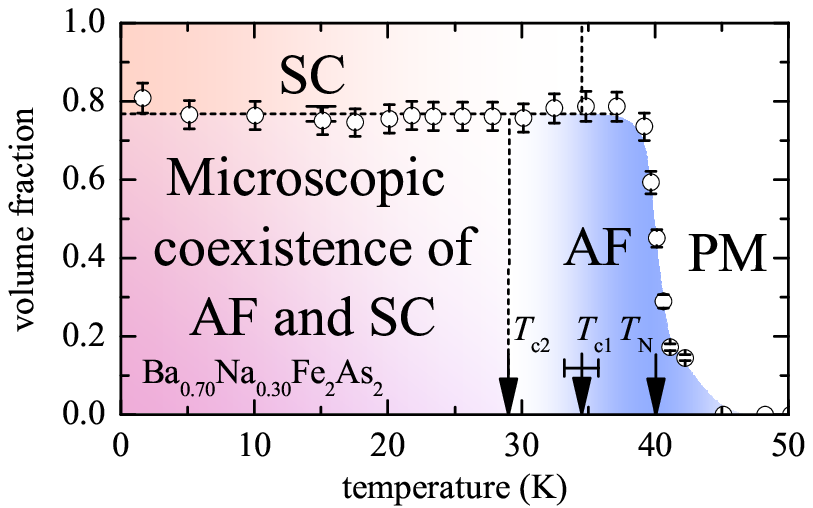} \label{img.phases}}
\caption{(a) We use TF $\mu$SR to estimate the paramagnetic volume fraction of \co{} for $T_{c2}<T<T_N$ to scale the ZF precession amplitude and obtain the magnetic volume fraction shown as circles in (b), the phase diagram of the \co{} sample: Below $T_N=40.0(5)$~K, 77(2)\% of the sample volume developed long range antiferromagnetic order.
Below $T_{c2}=29.0(6)$~K, the same volume also became superconducting, which was accompanied by a reduction of the ordered magnetic moment by $\approx 65$\%.
The remaining 23(2)\% sample volume stayed paramagnetic and became superconducting below $T_{c1}=34.5(13)$~K.}
\label{img.35-tf}
\end{figure}

\heading{Magnetic volume fraction} For $x=0.25$, the temperature independent total oscillation amplitude and TF-$\mu$SR show that 100\% of the sample is magnetically ordered (not shown).
For $x=0.30$, the remaining precession amplitude in 5~mT TF $\mu$SR indicates a paramagnetic phase with a volume fraction of $23(2)$\% below $T_N$, shown in Fig.~\ref{img.tf}.
The diamagnetic shift of the average internal field of the paramagnetic phase already for $T>T_{c2}=29.0(6)$~K indicates that this phase becomes superconducting below $T_{c1}=34.5(13)$~K.
The precession amplitude in zero field (see Fig.~\ref{img.phases}) is temperature independent, which indicates that the magnetic volume fraction reaches $\approx 77(2)$\% and remains temperature independent below $\approx 39$~K.
The sharp increase of the magnetic volume fraction in Fig.~\ref{img.phases} shows that, even though the transition temperature is reduced to $T_N=40.0(5)$~K, the transition remains sharp---much sharper than in other ferropnictides (measured from onset to saturation of the magnetic volume fraction it is not broader than 5(1)~K) \cite{wiesenmayer11,bernhard12,shiroka11,sanna09}.
The small damping rate $\lambda_T$ of the muon spin precession also indicates well defined magnetic order: it did not exceed $\lambda_T/(2\pi f_A)\approx 0.3$ for $x=0.30$ in the well-ordered magnetic state but also increased with doping, i.e., for $x=0.25$ $\lambda_T/(2\pi f_A)\approx 0.04$ at $T\approx 5 $~K (visible in Fig.~\ref{img.zf}).
The increase of the damping rate with doping indicates that Na doping indirectly causes disorder in the FeAs layers, albeit much weaker than Co doping that causes overdamped muon spin precession close to optimal doping \cite{bernhard12}.
Because Na is introduced between the FeAs layers, whereas Co resides directly in the magnetic FeAs layer, such behavior be expected.
We summarize the results of TF- and ZF-$\mu$SR for $x=0.30$ in a phase diagram shown in Fig.~\ref{img.phases}.

\heading{Rigid magnetic structure} The temperature dependence of the precession frequencies $f_A(T)$, $f_B(T)$ shown in Figs.~\ref{img.25-freq}, \subref*{img.35-freq} are identical and the temperature dependencies of the precession amplitudes, shown in Fig.~\ref{img.phases} for $x=0.30$ (scaled to the magnetic volume fraction), are constant below $\approx 0.95T_N$ for both magnetic compounds.
This observation indicates a magnetic structure that remains unaffected by superconductivity.
In particular, it indicates that the reduction of the precession frequencies for $x=0.30$ below $T_{c2}$ is solely due to the reduction of the ordered moment and not due to a reduction of the magnetic volume.
With a local probe like $\mu$SR this distinction is possible, whereas scattering techniques are only sensitive to the product of both quantities.

\begin{figure}[htb]
\includegraphics[width=8.5cm]{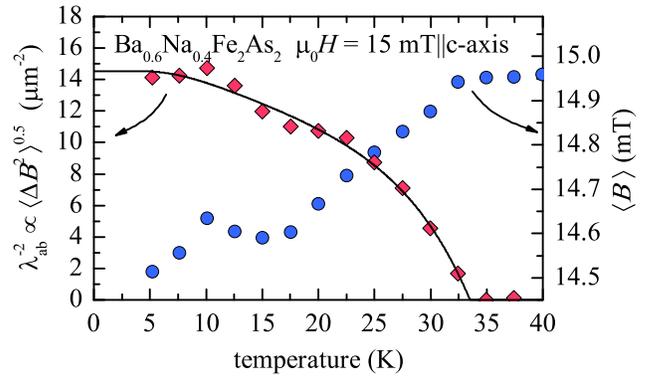}
\caption[]{The in-plane penetration depth $\lambda_{ab}^{-2}(T)$ (proportional to the superfluid density) and the (diamagnetic) average internal field of \od{} determined by TF $\mu$SR.}
\label{img.40}
\end{figure}

\heading{London penetration depth} The magnetic order for $x=0.25$, and 0.30 prohibits the study of the penetration depth by $\mu$SR.
For $x=0.40$ we measure the probability distribution $P(B)$ of local magnetic fields in the vortex state by TF-$\mu$SR with $\mu_0H=15$~mT parallel to the crystallographic $c$-axis.
Fig.~\ref{img.40} shows $\lambda^{-2}_{ab}(T)$, which is proportional to the standard deviation of $P(B)$, and the average local magnetic field \cite{brandt88}.
We find that $\lambda^{-2}_{ab}(T)$ is best described by two independent s-wave gaps without nodes $\Delta(0)=8.3(5)$, 3.0(4)~meV, $T_c=33.5(1)$~K, the weight of the large gap is $\omega=0.66(6)$, and $1-\omega$ for the small gap and $\lambda_{ab}(0)\approx 262$~nm (see supplemental material for details).
Vortex lattice disorder would artificially reduce $\lambda_{ab}$ (broaden $P(B)$), therefore $\lambda_{ab}(0)\approx 272$~nm is a lower limit for the in-plane penetration depth.
These values are in good agreement with $\Delta=10.5$, 3~meV determined ARPES on a single crystal from the same batch with the same composition \cite{aswartham12}.
Specific heat measurements \cite{pramanik11} on a crystal with 35\% Na doping and $T_c=29.5$~K revealed $\Delta=5.27$, 2.68~meV, which indicates that the small gap increases linearly with $T_c$ ($2\Delta/(k_BT_c)=2.1$ for both compositions), whereas the large gap grows non-linear with $T_c$ ($2\Delta/(k_BT_c)=4.2, 5.8(4)$ for $x=0.35$, 0.40, respectively).

\heading{Summary}
In summary, we have conducted muon-spin relaxation experiments on \ba{}, with $x=0.25$, 0.30, and 0.40.
The Na doping reduces the magnetic ordering temperature to $T_N=123(1)$, and 40.0(5)~K for $x=0.25$, and 0.30, respectively, but the magnetic transitions remains sharp and the magnetic order well defined.
For $x=0.40$, magnetic order and slow fluctuations were absent down to $T\approx 2$~K, as opposed to optimally doped BaFe$_{2-x}$Co$_x$As$_2$ \cite{bernhard12}.
All three specimens are bulk superconductors with critical temperatures $T_c=9$, 34.5(13), and 33.5(2)~K for $x=0.25$, 0.30, and 0.40, respectively ($T_c$ for $x=0.25$ from Ref.~\cite{aswartham12}).
For $x=0.25$, and 0.30 we presented unambiguous evidence for microscopic coexistence of the two phases.
The reduction of the magnetic order parameter by superconductivity was strongly dependent on doping: it was too weak to detect for $x=0.25$, but amounted to $\approx 65$\% for $x=0.30$.
For underdoped \bk{}, on the other hand, \citet{wiesenmayer11} reported a much smaller reduction of the ordered magnetic moment below $T_c$ that did not increase significantly with doping.
Within our itinerant model this can be accounted for by a much weaker superconducting order parameter relative to the magnetic one, i.e., better nesting of the electron and hole bands.
On the other hand, the non-monotonic doping dependence points to disorder effects which could suppress magnetism and superconductivity differently \cite{fernandes12}.
The presence of a second, purely superconducting minority phase for $x=0.30$ suggests chemical inhomogeneity---nevertheless, microscopic coexistence occurred in the majority of the sample volume (77(2)\%), suggesting that this phenomenon is more robust against disorder and doping in this system than in \bk.
The symmetry and size of the superconducting gaps, as well as $T_c$ for $x=0.40$ is similar to those of \bk{} \cite{khasanov09b}.
Considering the evidence for microscopic coexistence presented by \citet{wiesenmayer11} in underdoped \bk{} polycrystals and in this Letter for under- and nearly optimally doped \ba{}, we conclude that this phenomenon is intrinsic to hole doped BaFe$_2$As$_2$ and can be described by an itinerant two-band model as for BaFe$_{2-x}$Co$_x$As$_2$ \cite{fernandes10-order}.
\begin{acknowledgments}
HM thanks R. Fernandes for helpful discussions.
This work was funded by the German Research Foundation (DFG) within the priority program SPP1458, and the graduate school GRK1621.
JK acknowledges support from the German National Merit Foundation and the IMPRS Dynamical Processes in Atoms, Molecules and Solids.
SW acknowledges support by the DFG under the Emmy-Noether program (Grant No. WU595/3-1).
Part of this work has been performed at the Swiss Muon Source at the Paul Scherrer Institute, Switzerland.
\end{acknowledgments}


%

\clearpage
\pagebreak[4]

\appendix
\section{Analysis of $\lambda^{-2}_{ab}(T)$}
In analogy to Ref.~\cite{khasanov09b} we analyzed $\lambda_{ab}^{-2}(T)$ assuming two independent contributions to the superfluid density, arising from two $s$-wave gaps without nodes with different gap sizes $\Delta_1(0)$ and $\Delta_2(0)$ but identical $T_c$.
We let the gap size independent of $T_c$, which is the so called $\alpha$-model \cite{alphamodel}.
We fit the following expression to the data \cite{carrington03,tinkham2004}
\begin{equation}
\begin{split}
\lambda_{ab}(T)^{-2}=&\lambda_{ab}(0)^{-2}\left[\omega (1-D(\Delta_1(T),T))\right.\\
&\left.+(1-\omega) (1-D(\Delta_2(T),T))\right],\\
D(\Delta(T),T)=&2\int_\Delta^\infty -\frac{\partial f(E)}{\partial E} \frac{E}{\sqrt{E^2-\Delta(T)^2}}\,{\textrm d} E,
\end{split}
\end{equation}
where $f(E)$ is the Fermi-Dirac distribution, $\omega$ the weighting factor of the gaps, and the temperature dependence $\Delta(T)$ given by \citet{carrington03}.
\end{document}